# Impact of electron-phonon interactions on quantum-dot cavity quantum electrodynamics


Y. Ota[1, 2*], S. Iwamoto[1, 2], N. Kumagai[1] & Y. Arakawa[1, 2]

[1]Institute for Nano Quantum Information Electronics, The University of Tokyo, 4-6-1 Komaba, Meguro, Tokyo 153-8505, Japan

[2]Institute of Industrial Science, The University of Tokyo, 4-6-1 Komaba, Meguro, Tokyo 153-8505, Japan




Semiconductor quantum dots (QDs) in photonic nanocavities[1-8] provide monolithic, robust platforms for both quantum information processing[9,10] and cavity quantum electrodynamics (QED)[11,12]. An inherent feature of such solid-state cavity QED systems is the presence of electron-phonon interactions[13-18], which distinguishes these systems from conventional atomic cavity QED[11,12]. Understanding the effects of electron-phonon interactions on these systems is indispensable for controlling and exploiting the rich physics that they exhibit. Here we investigate the effects of electron-phonon interactions on a QD-based cavity QED system. When the QD and the cavity are off-resonance, we observe phonon-assisted cavity mode emission that strongly depends on the temperature and cavity-detuning. When they are on-resonance, we observe an asymmetric vacuum Rabi doublet, the splitting of which narrows with increasing temperature. These experimental observations can be well reproduced using a cavity QED model that includes electron-acoustic-phonon interactions[16]. Our work provides significant insight into the important but hitherto poorly understood mechanism of non-resonant QD-cavity coupling[3-7,18-21] and into the physics of various cavity QED systems



**utilizing emitters coupled to phonons, such as nitrogen-vacancy centres in diamond[22] and colloidal nanocrystals[23].**

Self-assembled semiconductor QDs are often referred to as artificial atoms due to their quantized energy levels for confined electron-hole pairs. However, the QDs are embedded in a crystalline host material and are intrinsically different from single, isolated atoms because they are subject to strong interactions with their surroundings, especially with phonons. For example, it is well known that the emission and absorption spectra of QDs are strongly affected by their coupling to phonons; the spectra consequently exhibit sidebands and/or Stokes lines in addition to the zero-phonon line[13-15]. Ever since the early stages of research into QD-based cavity QED systems, it has been recognised that phonons have a profound influence on their physics[2,16]. Wilson-Rae and Imamoglu[16] have theoretically investigated the effects of phonons and predicted the presence of spectral features that are unexpected in conventional cavity QED: an asymmetric vacuum Rabi doublet and narrowing of its splitting with increasing temperature. These phenomena have also been discussed by Milde *et al*[17], but neither has yet been experimentally observed. Phonon-mediated off-resonant QD-cavity coupling has also been discussed[4,18],



stimulated by recent experimental observations of bright off-resonant cavity mode emission[3-7]; this is known as the non-resonant QD-cavity coupling effect. Tarel and Savona[18] have studied these phonon-assisted coupling mechanisms theoretically, but the effects have not yet been experimentally proven. In this letter, we report a comprehensive study of the above issues. Our results demonstrate the striking effects of an electron-acoustic-phonon interaction on a QD-based cavity QED system.

We have investigated a system consisting of a two-dimensional, air-bridged H1 photonic crystal nanocavity[8] coupled with self-assembled InAs QDs. A scanning electron micrograph of the fabricated cavity is shown in Fig. 1a. The cavity quality factor is $Q$ = 11400 (the corresponding linewidth is 116 μeV). The areal density of the QDs is approximately 4 per μm$^2$ and the average number inside the cavity is estimated to be 0.4. Details of the device design, fabrication process and optical characterization techniques are given in the Methods section.

Figure 1b shows a photoluminescence (PL) spectrum of the system measured at 10 K. In addition to the QD-exciton emission peak at $E_{QD}$ = 1.33258 eV, a prominent cavity mode emission peak at $E_C$ = 1.33338 eV is apparent despite



the large cavity detuning of $\delta = E_C - E_{QD} = 810$ μeV. The single-emitter nature of the

system was confirmed by measuring the intensity auto-/cross-correlation

functions[3,24] $g^{(2)}(t) = \langle : I_i(t+\tau) I_j(t) : \rangle / \langle I_i(t) \rangle \langle I_j(t) \rangle$, where $::$ denotes the

normal ordering and $I_k(t)$ is the intensity of the emitted photons in the $k$-mode ($k =$

QD or cavity) at time $t$. The measured $g^{(2)}(t)$ functions for the QD and cavity

emission lines are shown in Figs. 1c-e. All the experimental data display

anti-bunching ($g^{(2)}(0) < 1$), which confirms that the dominant contribution to the

emission in this system is from a single QD[24]. This is a demonstration of

non-resonant QD-cavity coupling. We observed anti-bunching under all

experimental conditions that we examined: for zero and negative detuning and also

at low temperature (3.3 K).

Firstly, we discuss the dependence of PL from the system on the

cavity-detuning $\delta$ at 3.3 K and 10 K. The emission energy of the cavity mode was

scanned across the QD emission line using the nitrogen gas deposition technique[25]

at each fixed temperature. Figures 2a and b show colour plots of the obtained PL

spectra as a function of cavity detuning. In both plots anti-crossing behaviour, a

signature of the strong coupling regime[1,2], can be clearly seen. At 3.3 K, although



QD-like emission dominates the PL, a change in the intensity of the cavity-like emission is observed as the sign of the detuning changes. Although such asymmetry in the non-resonant coupling has previously been reported in the literature (for example, see ref. 3), the origin of this feature has not been identified. Interestingly, the intensity ratio of the cavity-like emission to the QD-like emission increases at large absolute cavity detuning $|\delta|$ (see also Figs. 2f-g). These phenomena cannot be explained solely by the mechanisms that have previously been used to describe the non-resonant coupling, the photon-mediated shake-up process[5] and pure-emitter-dephasing[19-21]. At 10 K, the PL properties change drastically. The spectra become more symmetric with respect to the sign of cavity detuning. Moreover, the cavity-like emission at large detuning ($|\delta| > 0.4$ meV) is considerably stronger than the QD-like emission.

In order to understand the origin of these observations, we performed numerical simulations based on a master equation method including the electron-acoustic-phonon interaction. The model used is described in Fig. 2e. A single cavity mode is coupled with a QD described by a two-level-emitter, where the upper level is coupled to a longitudinal-acoustic (LA) phonon reservoir. This is



an extension of the independent boson model[15] and is in principle equivalent to models used in previous work[16-18]. We followed the treatment described by Wilson-Rae and Imamoglu[16] and arrived at the following integro-differential master equation for the reduced density matrix $\rho$(t) under the second-order Born approximation:

$$\frac{\partial \boldsymbol{\rho}(\boldsymbol{t})}{\partial \boldsymbol{t}} = -\frac{\boldsymbol{i}}{\hbar}\left[\boldsymbol{H_{sys}}, \boldsymbol{\rho}(\boldsymbol{t})\right] + \boldsymbol{L}\boldsymbol{\rho}(\boldsymbol{t}) - \frac{1}{\hbar^2}\int\limits_0^t \boldsymbol{\Lambda}(\boldsymbol{\tau})\boldsymbol{\rho}(\boldsymbol{t}-\boldsymbol{\tau})\boldsymbol{d}\boldsymbol{\tau}, \tag{1}$$

where $H_{sys}$ is the system Hamiltonian including time-independent phonon effects and $L$ is a Liouvillian describing Markovian processes that include cavity loss, spontaneous emission, incoherent QD pumping and QD pure dephasing. The last term expresses the time-dependent interaction with the non-Markovian phonon reservoir, whose effects are described by the super-operator $\Lambda(\tau)$. Emission spectra for the QD-cavity system can be obtained from equation (1) by using the quantum regression theorem under the Born approximation and by the Wiener-Khinchine theorem[19]. The main parameters used in the simulations are summarized in the table in Fig. 2e. Details of the calculation method are given in the Supplementary Information.

The spectra calculated for temperatures of 3.3 K and 10 K are shown in



Figs. 2c and d. At 3.3 K, both the asymmetric feature in the non-resonant QD-cavity coupling and the bright cavity emission at large absolute cavity detuning are successfully reproduced. In addition, the more symmetric and stronger non-resonant QD-cavity coupling at 10 K is also reproduced. For further analysis, we plot the relative intensities of the QD-like and cavity-like emission branches as a function of cavity detuning in Figs. 2f-i. A general feature of all these curves is that the QD-like emission becomes stronger near the QD-cavity resonance, $\delta \sim 0$. This behaviour indicates the dominant contribution of Purcell-effect enhanced[26] emission from the QD zero-phonon line around the resonance. In contrast, at larger detuning the QD-like emission decreases and the total emission is dominated by the phonon-mediated cavity mode emission. In addition, the intensity of the cavity-like branch is asymmetric with respect to the sign of the detuning; this feature is most prominent at 3.3 K and is attributed to an imbalance in the phonon emission and absorption processes. The QD rarely absorbs phonons at low temperature, which results in weak coupling of the QD to the positively detuned ($\delta > 0$) cavity mode. Thus, the off-resonant cavity mode emission at $\delta > 0$ becomes weaker than for detuning of opposite sign, where the phonon-emission process can more frequently



mediate between the QD and the cavity. When the temperature is increased to 10 K, the phonon emission and absorption processes become more active and more balanced. Thus, the cavity-like branch shows stronger emission than at 3.3 K and is more symmetric with respect to the sign of the detuning. Our calculations are in good agreement with these experimental features. We thus conclude that, in the detuning range investigated ($|\delta| < 1$ meV), the LA-phonon-electron interaction plays a dominant role in the observed non-resonant QD-cavity coupling. The remaining disagreement between calculation and experiment might be due to other types of electron-phonon coupling[13,15,16].

We will now discuss the on-resonance emission properties of our system ($\delta = 0$) with respect to temperature. Figures 3a-c show the measured and calculated PL spectra containing the vacuum Rabi doublet at three different temperatures. The doublet emission exhibits asymmetry, the degree of which is reduced with increasing temperature. The origin of this asymmetry can again be attributed to the imbalance in the phonon emission and absorption processes across the zero-phonon line[17]. We also observe that the vacuum Rabi splitting decreases monotonically with increasing temperature, as shown in Fig. 3d. This phenomenon is mainly due to a



phonon-induced effective reduction of the QD-cavity coupling strength[16] $g$ by a factor of $<B>$ ($<B> \leq 1$, $<B>$ decreases towards zero as the temperature increases). These observations demonstrate that electron-phonon interaction adds asymmetry to the vacuum Rabi doublet and suppresses the effective QD-cavity coupling strength.

In conclusion, we have observed the large influence of electron-phonon interactions on a QD-based cavity QED system. In particular, we have demonstrated that the LA-phonon-electron interaction plays a significant role in the non-resonant QD-cavity coupling mechanism. Moreover, we have experimentally observed the asymmetry in the vacuum Rabi doublet and the decrease of its splitting with increasing temperature that were previously predicted in the literature. These experimental observations can be consistently reproduced using a cavity QED model that includes the interaction between the QD and the LA phonons. The model can easily be applied to various other forms of electron-phonon interactions. Thus, we believe that this work will pave the way for studies of electron-phonon interactions in many cavity QED systems. One interesting future direction of this research is the exploitation of phonon-mediated non-resonant QD-cavity coupling



effects in emission-energy-stabilized single photon sources[27-29].



**METHODS**

**Sample growth and preparation**. The semiconductor heterostructure used for the fabrication of the photonic crystal nanocavity was grown on a (100)-oriented semi-insulating GaAs substrate by molecular beam epitaxy. The wafer consisted of a 160-nm-thick GaAs slab incorporating a layer of self-assembled InAs QDs at the centre, deposited on top of a 700-nm-thick $Al_{0.6}Ga_{0.4}As$ sacrificial layer. The partially covered island growth technique[30] was applied to the QDs in order to obtain blue-shifted emission at ~1.333 eV (measured at 6 K). The cavity design is a so-called symmetrically-modified H1, the details of which are described in ref. 8. The cavity was fabricated by a combination of electron beam lithography and dry/wet etching processes. We focused on one of the fundamental dipole-like modes, which are theoretically doubly degenerate but in practice are split due to fabrication errors. The other mode with orthogonal polarization was found at 7 meV below the investigated mode and was excluded from discussion in this Letter.

**Optical characterization.** We characterized the QD-cavity system using micro-PL measurements. The sample was mounted in a temperature-controlled liquid helium cryostat. The excitation source was a continuous-wave Ti:sapphire laser oscillating



at 840nm. The excitation laser was focused onto the sample surface with a spot size of ~3μm using a microscope objective lens (40x, N.A. = 0.6). The PL signal was collected using the same objective lens and sent to a 0.75 m grating spectrometer equipped with a cooled CCD camera (spectral resolution = 23μeV) after passing through a half-wave plate and a linear polarizer. For photon auto-/cross-correlation measurements, a Hanbury-Brown-Twiss setup consisting of a pair of single photon counters located after monochromators functioning as bandpass filters (bandwidth 0.36 meV) was employed. The detector time resolution was approximately 400 ps. The recorded intensity correlation functions were fitted using the function[24,29]

$$g^{(2)}(t) = 1 - (1 - g^{(2)}(0)) \exp(-|t|/t_m),$$ where $t_m$ is the anti-bunching time constant. In order to control the emission energy of the cavity mode, we employed the nitrogen gas deposition technique[25]. The cavity was scanned continuously but slowly enough (at an average rate of 4 μeV per second) to obtain quasi-static spectra within the accumulation time of the CCD camera (0.5 s). The recorded spectra were analyzed by fitting with two Lorentzian curves. The spectra at the resonance condition shown in Fig. 3 were extracted from those with the smallest splitting among a series of cavity-scanned spectra.



**Figures**

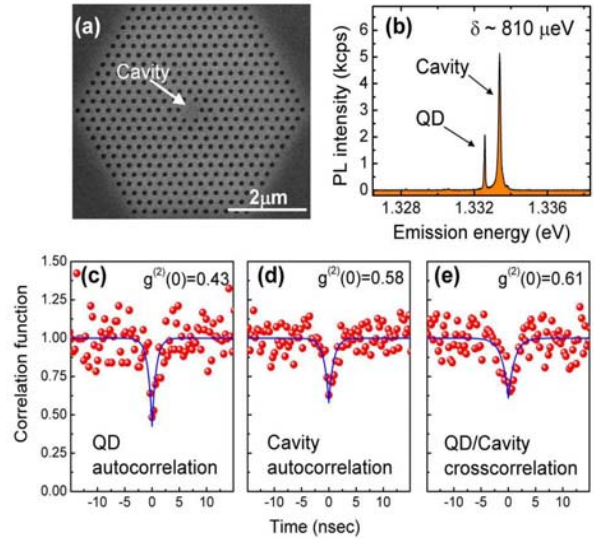

**Figure 1** Photonic crystal structure and optical characteristics. **a**, Scanning electron micrograph of the H1 photonic crystal nanocavity investigated. **b**, PL spectrum of the coupled QD-cavity system at 10 K for a detuning of $\delta \sim$ 810 µeV. The excitation power was 0.5 µW (measured before the objective lens). The QD-exciton and cavity emission peaks can be clearly seen. **c-e,** Measured intensity correlation functions $g^{(2)}(t)$ of QD auto-correlation (**c**), cavity auto-correlation (**d**), and QD-cavity cross-correlation (**e**). The data were taken under the same conditions as for **b**. The red spheres and blue lines denote experimental data and fits, respectively.



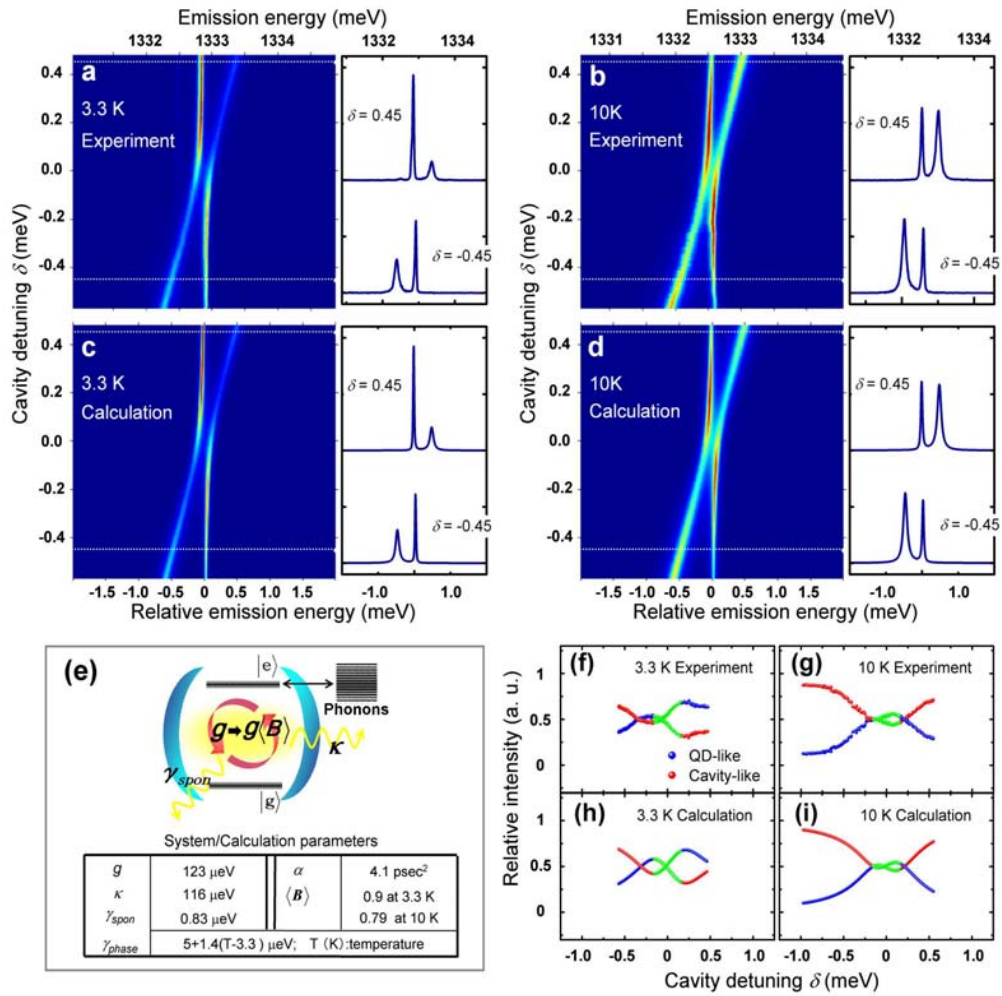

**Figure 2** Effects of detuning and temperature on PL from the QD-cavity system. **a-b**, Colour plots of PL spectra measured at various cavity detunings $\delta$ at 3.3 K (**a**) and 10 K (**b**). The intensity ranges from zero (blue) to one (red), going through light blue, green and yellow. The right-hand panels show the spectra for detuning $\delta$ = ±0.45 meV, the locations of which in the colour plots are indicated by white dotted lines. The excitation power



was 0.15 µW. The cavity-like emission is assigned to the lower energy branch when $\delta$ < 0 and to the higher energy branch when $\delta$ > 0 (vice versa for QD-like emission) **c-d**, Calculated spectra corresponding to **a-b**. In **a-d**, every spectrum at a specific $\delta$ was normalized to its total intensity. **e**, Cavity QED model including electron-phonon interactions; $|e\rangle$ and $|g\rangle$ denote the excited and ground states of the QD, respectively. The QD interacts with the cavity field with coupling strength $g$ and also couples with phonons (coupling parameter $\alpha$). The effective QD-cavity coupling strength is reduced to $g<B>$ through the QD-phonon interaction. The parameters $\kappa$ and $\gamma_{spon}$ denote the cavity leakage and spontaneous emission decay, respectively. Weak incoherent QD pumping $P_{QD}$ = 0.08 µeV ($P_{QD}$ /2$\pi$ = 0.02 GHz) from $|g\rangle$ to $|e\rangle$ and temperature-dependent QD pure dephasing ($\gamma_{phase}$) are also considered in the model. The system/calculation parameters are summarised in the table. **f-g**, Measured PL integrated intensities of QD-like and cavity-like emission relative to the total intensity, plotted as a function of $\delta$ at 3.3 K (**f**) and 10 K (**g**). **h-i**, Calculated PL intensity corresponding to **f-g**. In **f-i**, the blue (red) spheres correspond to QD-like (cavity-like) emission.



The green spheres denote the region $|\delta| < 0.15$ meV, where the QD-like and cavity-like emission are largely mixed.

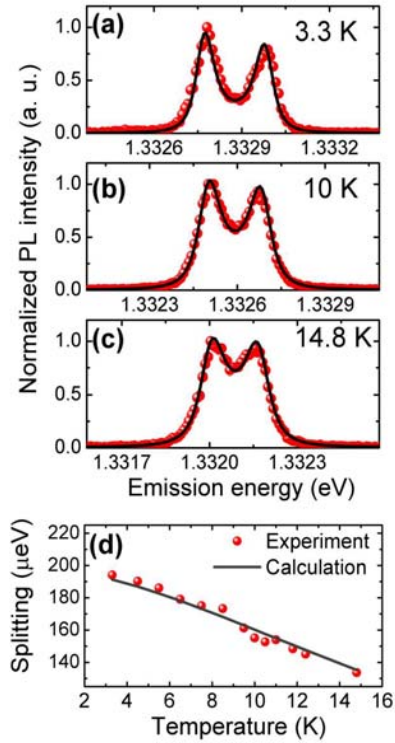

**Figure 3** Asymmetric vacuum Rabi doublet and the decrease in splitting with increasing temperature. **a-c.** Measured and calculated PL spectra at 3.3 K (**a**), 10 K (**b**) and 14.8 K (**c**). The excitation power was 0.15 μW. As the temperature increases, the spectra become more symmetric. **d.** Vacuum Rabi splitting energy as a function of temperature. In **a-d**, the red spheres indicate measured data and the black curves represent calculated results.



## Acknowledgements


We thank S. Yorozu, S. Ohkouchi, M. Shirane, Y. Igarashi, S. Ishida, M. Nomura and K. Watanabe for their technical support and for fruitful discussions. Y. O. thanks T. Nakaoka, T. Miyazawa and S. Kako for their comments on the paper. This research was supported by the Special Coordination Funds for Promoting Science and Technology, Japan.


## Competing Interests

The authors declare that they have no competing financial interests.

## Correspondence

Correspondence and requests for materials should be addressed to Y. O. (e-mail: ota@iis.u-tokyo.ac.jp) or Y. A. (e-mail: arakawa@iis.u-tokyo.ac.jp).

# Supplementary Information

## Numerical simulations:

We simulated the emission spectra from the strongly coupled quantum-dot (QD) and cavity system using a quantum master equation method that includes interaction with longitudinal-acoustic (LA) phonons. The model is schematically described in Fig. 2e in the main text and the mathematical derivation is based on the procedures described in ref. 1. First, the system is projected onto the three states in the Hilbert space spanned by $|g, n_c = 0\rangle$ ( $|0\rangle$ ), $|g, n_c = 1\rangle$ ($|1\rangle$), $|e, n_c = 0\rangle$ ($|2\rangle$) by imposing an upper limit of 1 on the cavity photon number $n_c$. This approximation is adequate for our situation where the system is weakly excited and is effective in saving calculation time. The system Hamiltonian in the rotating frame at the QD resonant frequency is then expressed as; $H_{sys} = \delta\sigma_{11} + \hbar g\langle B\rangle(\sigma_{21} + \sigma_{12})$, where $\sigma_{ij} = |i\rangle\langle j|$ is a pseudo-Pauli spin operator, $\delta$ is the detuning of the cavity from the QD resonance and $g\langle B\rangle$ is the effective QD-cavity coupling strength when the electron-phonon interaction occurs. The master equation (1) in the main text can be derived by dealing with the non-Markovian phonon reservoir in a thermal state at the lattice temperature and by adding other Markovian processes at zero temperature. The Markovian processes are described as follows:

$$
\begin{aligned}
L\rho &= \frac{\kappa}{2}(2\sigma_{01}\rho\sigma_{10} - \sigma_{11}\rho - \rho\sigma_{11}) + \frac{\gamma_{spon}}{2}(2\sigma_{02}\rho\sigma_{20} - \sigma_{22}\rho - \rho\sigma_{22}) \\
&+ \frac{P_{QD}}{2}(2\sigma_{20}\rho\sigma_{02} - \sigma_{00}\rho - \rho\sigma_{00}) + \frac{\gamma_{phase}}{2}((\sigma_{22} - \sigma_{00})\rho(\sigma_{22} - \sigma_{00}) - \rho)
\end{aligned}
\tag{S1}
$$

These four terms represent the cavity loss ($\kappa$), spontaneous emission ($\gamma_{spon}$), incoherent QD pumping ($P_{QD}$), and QD pure dephasing ($\gamma_{phase}$), respectively. The parameter $\gamma_{phase}$ is

introduced to obtain better agreement between the calculations and experiments. However, the most interesting features mainly originate from the phonon effects, a conclusion that is confirmed by performing calculations with $\gamma_{phase} = 0$. For the temperature dependence of $\gamma_{phase}$, we used the value of 1.4 μeV/K measured for the zero-phonon linewidth of InAs/GaAs QDs embedded in mesa structures[2] with approximate diameters of 500 nm, which is the same order as the geometric defect size of our cavity (~ 500 nm).

For the calculation of the spectra in the steady-state at time $\tau$, we first computed two-time correlation functions using the quantum regression theorem[3] with the expectation values of the operators $\left\langle \sigma_{ij}(\tau) \right\rangle$. The spectra of the cavity leak channel $S_{CAV}(\omega)$ and of the QD direct emission channel $S_{QD}(\omega)$ were then calculated using the Wiener-Khintchine theorem:

$$S_{CAV}(\omega) = \frac{\kappa}{2\pi} \left\{ \int_0^\infty \left\langle \sigma_{10}(t+\tau)\sigma_{01}(\tau) \right\rangle e^{-i\omega t} dt + \int_0^\infty \left\langle \sigma_{10}(\tau)\sigma_{01}(t+\tau) \right\rangle e^{i\omega t} dt \right\},$$

$$S_{QD}(\omega) = \frac{\gamma_{spon}}{2\pi} \left\langle B \right\rangle^2 \left\{ \int_0^\infty \left\langle \sigma_{20}(t+\tau)\sigma_{02}(\tau) \right\rangle e^{\phi(t)-i\omega t} dt + \int_0^\infty \left\langle \sigma_{20}(\tau)\sigma_{02}(t+\tau) \right\rangle e^{\phi(t)^*+i\omega t} dt \right\}, \qquad (S2)$$

where $\phi(t) = \int_0^\infty \frac{J(\omega)}{\omega^2} \left[ \coth(\frac{\beta\hbar\omega}{2})\cos(\omega t) - i\sin(\omega t) \right] d\omega$.

The quantity $\beta$ is the inverse of the product of the Boltzmann constant and the temperature. The parameter $J(\omega)$ is the spectral function with the simplified form[4] $(4\pi^2)^{-1}\alpha\omega^3 \exp(-\omega^2/2\omega_b^2)$, describing the electron-LA-phonon interaction (deformation potential coupling). We set $\alpha$, the coupling parameter between the electron and the LA-phonons, to a value of 4.1 ps[2]. This is 20 times larger than the electron-LA-phonon coupling parameter in bulk GaAs, which can be expressed as $(D_C - D_V)^2/\rho u^5$ and can be calculated from the sound velocity of the longitudinal acoustic wave ($u$ = 5110 m/sec), the density ($\rho$ =

5370 kg/m$^3$), and the deformation potentials for the valence ($D_V$ = -4.8 eV) and conduction ($D_C$ = -14.6 eV) bands; all of these parameters were taken from ref. 5. The multiplication by a factor of 20 was performed to obtain better quantitative agreement between the calculated and the experimental results. This additional factor does not affect the basic results qualitatively. The parameter $\omega_b$ is the cut-off frequency and corresponds to the inverse of the flight time of the phonon through the QD. We set $\omega_b$ to 0.84 meV. The total spectrum at the detector is expressed as $\eta_{CAV} S_{CAV}(\omega) + \eta_{QD} S_{QD}(\omega)$, where $\eta_{CAV/QD}$ are the coupling efficiencies of each channel to the detector. Here we set $\eta_{CAV/QD}$ = 1. The spectra were convolved with a Gaussian function with a full width at half maximum of 23 μeV (detector resolution). The spectra presented in the main text were normalised to their total intensities. For the colour plots in Fig. 2, we scanned the cavity energy in steps of 10 μeV. For extraction of the emission properties, the spectra were fitted using two Lorentzian curves.